\documentclass[prb,twocolumn,showpacs,preprintnumbers,amsmath,amssymb]{revtex4}
\usepackage{graphicx}% Include figure files
\usepackage{dcolumn}% Align table columns on decimal point
\usepackage{bm}% bold math

\begin{document}

\title{
  Effect of $A$-site size difference on polar behavior in $M$BiScNbO$_6$,
  ($M$$=$Na, K and Rb): Density functional calculations
}

\author{Shigeyuki Takagi$^{1,2}$}
\author{Alaska Subedi$^{1,2}$}
\author{Valentino R. Cooper$^1$}
\author{David J. Singh$^1$}
%\email{takagism@ornl.gov}
\affiliation{
  $^1$Materials Science and Technology Division,Oak Ridge National Laboratory,
  Oak Ridge, TN 37831-6114 \\
  $^2$Department of Physics, University of Tennessee, Knoxville,
  TN 37996
}

\date{\today}

\begin{abstract}
  We investigate the effect of $A$-site size differences in the double
  perovskites BiScO$_3$-$M$NbO$_3$ ($M$$=$Na, K and Rb) using
  first-principles calculations. We find that the polarization of these
  materials is 70$\sim$90 $\mu$C/cm$^2$ along the rhombohedral
  direction. The main contribution to the high polarization comes
  from large off-centerings of Bi ions, which are strongly enhanced by
  the suppression of octahedral tilts as the $M$ ion size increases.
  A high Born effective charge of Nb also contributes to the
  polarization and this contribution is also enhanced by
  increasing the $M$ ion size. 
\end{abstract}


\maketitle

\section{INTRODUCTION}
Ferroelectric
perovskite oxides have attracted attention for many years because of their
many applications in technology, including
memory devices, electronic components, piezoelectric sensors and actuators,
and pyroelectric devices. \cite{scott}
Besides the technological importance,
there is also a great scientific interest
in understanding the underlying physics.
This is stimulated by recent discoveries of entirely new
ferroelectrics particularly in thin film form. \cite{ahn,lee,dawber}
The key to these materials is competition between different
interactions, leading to structural instability.
Ferroelectricity in perovskite $AB\text{O}_3$ oxides depends on the
delicate balance of the long-range Coulomb energy, which favors a ferroelectric
distortion, and the short-range ionic repulsion, which stabilizes the cubic
phase.\cite{cohen1990prb,cohen1992nature}
More generally, instabilities in perovskites are often discussed in terms
of competition between preferred $A$-O and $B$-O bond lengths, which
cannot both be accommodated in the ideal cubic structure
for arbitrary $A$ and $B$ ions.
This is characterized by a so-called tolerance factor, $t$, that describes
the degree of non-satisfaction of ideal bond lengths in the cubic
perovskite structure.
The balance between the long-range Coulomb energy and short-range
ionic repulsion can be modified by smaller effects related to the
$A$-O hybridization, in particular the so-called lone pair
physics associated with stereochemically active ions, particularly
Pb$^{2+}$ and Bi$^{3+}$, as in the technologically important
Pb(Zr,Ti)O$_3$ system and BiFeO$_3$.
In thin films and superlattices
the constraint imposed by lattice matching imposes strains that
push bond lengths away from their ideal values, leading to highly enhanced
ferroelectric instabilities in many cases. \cite{lee,dawber,cooper}

The piezoelectric performance in perovskite ferroelectrics
is associated with polarization rotation 
near morphotropic phase boundaries (MPB).
\cite{park1997jap,noheda1999apl,bellaiche2000prl,fu2000nature,guo2000prl,noheda2001prl}
That is, in materials where two different ferroelectric states with different
lattice strains are close in energy,
an electric field may rotate the polarization
and therefore the ferroelectric state from the
zero field equilibrium configuration towards the other nearby
configuration, with resulting coupling to the lattice strain.
Strong ferroelectricity is highly desirable for piezoelectricity,
but epitaxial constraints on the lattice
strain as in most thin films are not.
It is of interest to examine strategies for producing
structural frustration and perhaps enhance
ferroelectricity in bulk materials.

In this regard, recent first principles calculations
suggest that $A$-site off-centering can be enhanced by the use
of ions with different ionic radii at the
$A$-site.\cite{singh2008prl} There is experimental support for this
in the (Ba,Ca)TiO$_3$ solid solution, which shows a strongly
stabilized ferroelectric tetragonal phase over a wide composition range
as compared to what would be expected from a classical
averaging of the end-point properties.\cite{desheng2008prl} 
However, while this is shown experimentally to be effective
in enhancing ferroelectricity in (Ba,Ca)TiO$_3$ it is not
effective in the closely related (Ba,Ca)ZrO$_3$ system.
\cite{bcz}
We recently
reported results of a first principles study of the
double perovskites, BiPbZnNbO$_6$ and BiSrZnNbO$_6$, which have 
a large size difference on the $A$-site.
\cite{takagi2010prb} We found high
polarization values of $\sim$80 $\mu$C/cm$^2$ in these compounds,
which is comparable to BiFeO$_3$ (90-100 $\mu$C/cm$^2$)
\cite{neaton2005prb} even though these double
perovskites have only half the $A$-sites
occupied by the stereochemically active Bi$^{3+}$
ion. The high polarization in
these materials results comes from large off-centerings of Bi ions,
associated in part with the $A$-site size mismatch,
cooperating with
moderate off-centerings of Nb ions that have Born effective
charges highly enhanced by covalency.

Here we use density functional calculations to
study the double perovskites $M$BiScNbO$_6$
  ($M$ $=$ Na, K and Rb) in order to
  further investigate the effect of $A$-site size differences on the
  polar behavior of perovskites.
In relation to BiSrZnNbO$_6$,
  the compounds that we study here have lower ionic charges at the
  $A$-site and higher ionic charges at the $B$-site. The $M$ ions have
  increasing ionic radii ($r_{\textrm{Na}^+} < r_{\textrm{K}+} <
  r_{\textrm{Rb}^+}$),
but are otherwise chemically similar, which is helpful in understanding how
  increasing $A$-site size difference affects the
  ferroelectricity.
Finally, we note that these materials could potentially be
of practical use.
It is known that
the end-point compound BiScO$_3$
  (BS) strongly favors a rhombohedral state over a tetragonal
  state, \cite{eitel,iniguez,halilov2004prb}
as does BiSrZnNbO$_6$.
As such these systems, when alloyed with a tetragonal
ferroelectric can show MPBs and possibly form useful piezoelectrics.
\cite{eitel}

\section{APPROACH}

  We investigated the lattice instabilities of three perovskite oxides
  (Na,Bi)(Sc,Nb)O$_6$, (K,Bi)(Sc,Nb)O$_6$ and (Rb,Bi)(Sc,Nb)O$_6$
  using density functional theory (DFT) within the local-density
  approximation (LDA). All calculations, including the structure
  relaxations, polarizations and Born effective charges, were done
  using a plane wave basis as implemented in Quantum Espresso
  package.\cite{espresso} We used ultrasoft
  pseudopotentials\cite{uspp} and converged our results with respect
  to the cut-off energy and Brillouin zone sampling, similar to our
  previous study for BiPbZnNbO$_6$ and
  BiSrZnNbO$_6$.\cite{takagi2010prb} The polarization was calculated
  using the Berry's phase method.

  There is a large charge difference between Sc$^{3+}$ and Nb$^{5+}$,
  and this is expected
lead to a strong ordering tendency on the $B$-site.
This is actually the case in the relaxor compound
  PbSc$_{1/2}$Nb$_{1/2}$O$_3$ (PSN), where Sc and Nb have rock-salt
  ordering.\cite{chu1995jap,malibert1997jpcond}
Here we consider material with this double perovskite type $B$-site ordering.
In addition, while ordering at the $A$-site is less clear,
there is a large charge difference at the $A$-site, which
may lead to an ordering tendency at least short range, as well.
There have been
  reports of short-range cation ordering of Na$^+$ or K$^+$ and
  Bi$^{3+}$ in Na$_{1/2}$Bi$_{1/2}$TiO$_3$ (NBT)
  and K$_{1/2}$Bi$_{1/2}$TiO$_3$
  (KBT),\cite{park1994jamceram,jones2002powder,jones2002acta,petzelt2004jphys}
  which support this view. Here we consider the ordered system, both because
it is simpler, which facilitates the calculations and analysis, and
because it helps to avoid computational artifacts, such as polarization
because of the particular
symmetry breaking chemical ordering in a given supercell.

  There are various factors at play that may drive ferroelectricity in
  NaBiScNbO$_6$ (NBSN), KBiScNbO$_6$ (KBSN) and RbBiScNbO$_6$
  (RBSN). These materials have Nb$^{5+}$ at the $B$-site which
  favors ferroelectricity, as in KNbO$_3$. The end-point
  perovskite BiScO$_3$ has a very small tolerance factor
  $t=0.85$ and a stereochemically active Bi$^{3+}$ at the $A$-site,
  two factors that are considered to be important in $A$-site driven
  materials, leading to strong lattice instabilities.
Furthermore, the large size
  difference at $A$-site may also enhance the off-centering of the
  $A$-site ions through frustration of tilt modes.
However, in addition to $A$-site off-centering, low
  tolerance factors also strongly favor
rotation of the $B$O$_6$ octahedra. These
rotations compete with ferroelectricity
in general.
In order to sort this out
it is important to use supercells that have even numbers of
  units along the [100], [010] and [001] directions to accommodate the
  various Glazer tilt patterns.\cite{glazer1972acta}
We used 40 atom 2$\times$2$\times$2 supercells and
  assumed both $A$- and $B$-site rocksalt chemical ordering.
  In addition to allowing various Glazer tilt patterns, these
  supercells are cubic and
therefore non-polar as regards to chemical ordering.
Thus, any off-centering is a consequence of lattice
  instabilities and not an artifact due to cation ordering in the
supercell.

\section{RESULTS AND DISCUSSION}

  To our knowledge, syntheses of NaBiScNbO$_6$, KBiScNbO$_6$ and
  RbBiScNbO$_6$ have not been reported.  We therefore used theoretical
  LDA lattice parameters.
We note that in oxides with heavy elements the LDA typically leads to
a small underestimation of the lattice parameter by $\sim$ 1-2\%.
If this is the case here, it will work against ferroelectricity
and in favor of competing tilt instabilities.
\cite{fornari-tilt}
We assume a pseudocubic structure to evaluate the lattice
  parameters and relaxed all atomic coordinates with no symmetry
  constraints.  This was done as a function of the lattice parameter
  to find the equilibrium volumes for these materials.  The calculated
  LDA lattice parameters for the minimum energy were 3.98 \AA, 4.03
  \AA\ and 4.07 \AA\ for NaBiScNbO$_6$, KBiScNbO$_6$ and
  RbBiScNbO$_6$, respectively.  These show the expected trend with
  increase in the $A$-site ion size.

\begin{figure}[ht!]
\begin{tabular}{cc}
\includegraphics[width=4.0cm,clip]{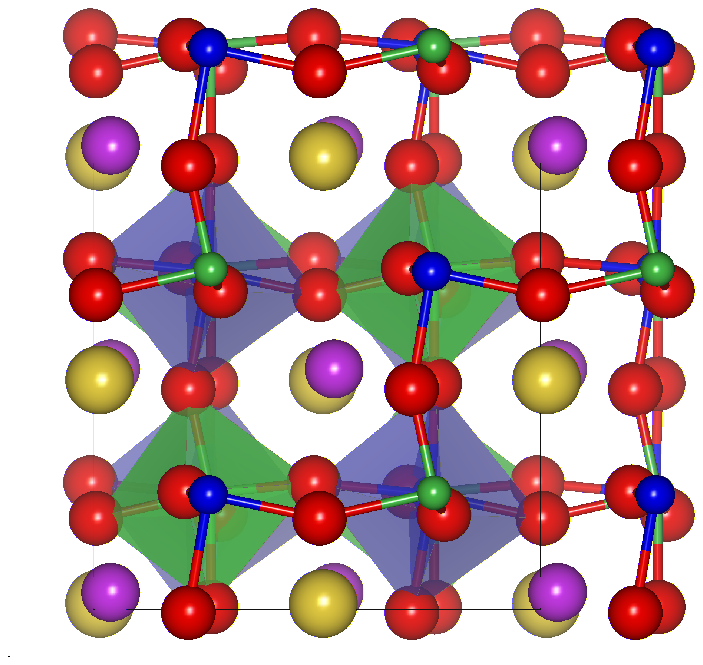} & \includegraphics[width=4.0cm,clip]{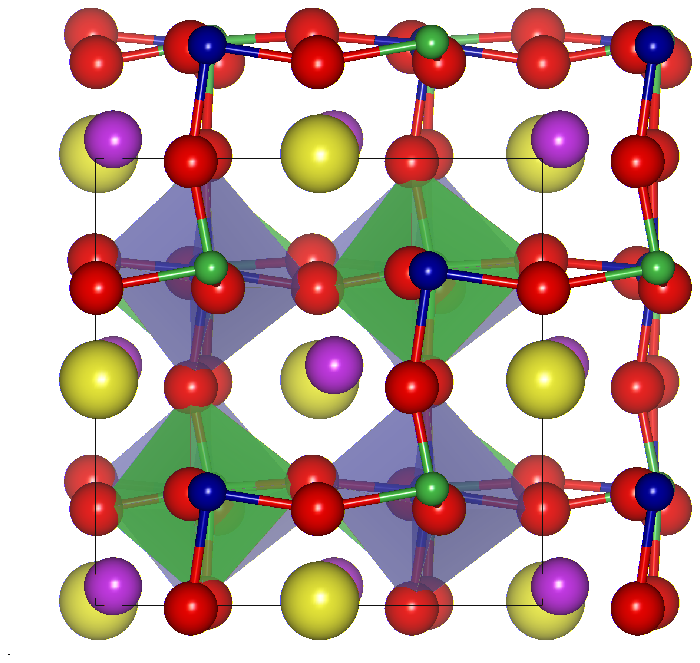} \\
(a) NaBiScNbO$_6$ & (b) KBiScNbO$_6$ \\
\multicolumn{2}{c}{\includegraphics[width=4.0cm,clip]{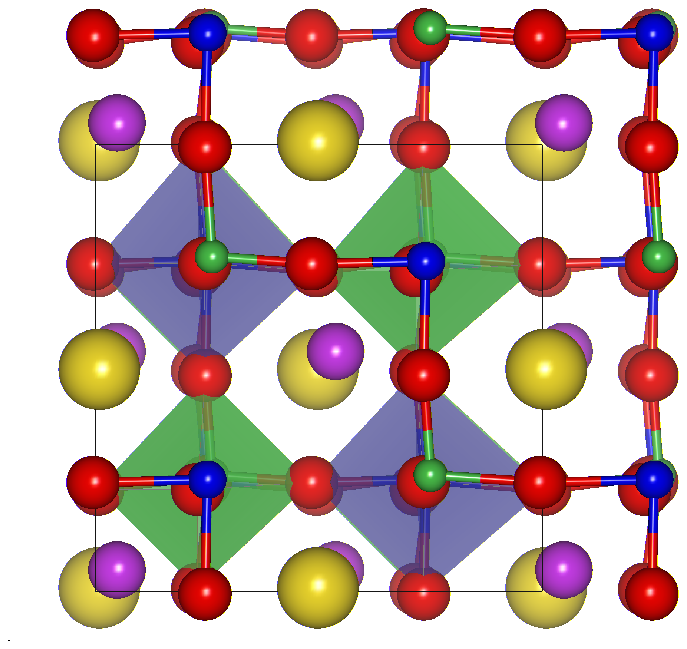}} \\
\multicolumn{2}{c}{(c) RbBiScNbO$_6$} \\
\end{tabular}
\caption{\label{structure}
  (color online) Projections onto a (010) plane of the structures of
  the relaxed 40 atom pseudocubic (a) NaBiScNbO$_6$, (b) KBiScNbO$_6$ and (c)
  RbBiScNbO$_6$ supercells.
  The $B$-O bonds are shown.
  The Bi ions are denoted by light purple spheres,
the Sc ions by blue, the Nb
  ions by light green, and the O ions by red.
  The Nb, K and Rb ions are shown by dark yellow spheres in each panel.
  The O octahedra for Sc are shown by (semi-transparent) blue boxes and those
  for Nb by (semi-transparent) green.
  The solid lines show the unit cell boundaries of each material.
}
\end{figure}

  Fig.~\ref{structure} shows the structures of the relaxed 40 atom
  pseudocubic NaBiScNbO$_6$, KBiScNbO$_6$ and RbBiScNbO$_6$ supercells
  at the equilibrium volume. As expected, the
  octahedral tilting is reduced with larger $A$-site ion size
  difference (note that the ionic radii of $A$-site ions are
  $r_{\text{Bi}^{3+}}=1.31$ \AA, $r_{\text{Na}^{+}}=1.53$ \AA,
  $r_{\text{K}^{+}}=1.78$ \AA\ and $r_{\text{Rb}^{+}}=1.86$ \AA,
  respectively). In particular, there is relatively
 little tilt in RbBiScNbO$_6$,
  which has the largest $A$-site size difference.
  Fig.~\ref{structure} also shows large displacements of the Bi
  ions from the ideal positions at the centers of the O cages.

\begin{figure}[ht!]
\includegraphics[width=8cm,clip]{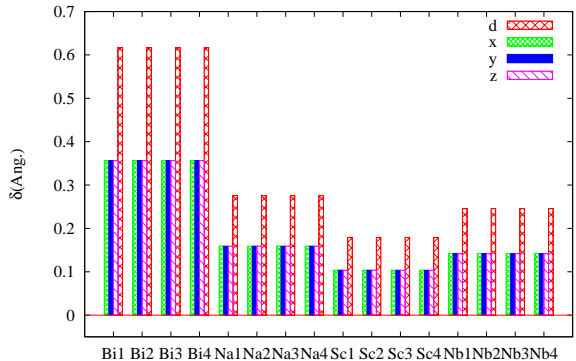}
\caption{\label{offcenter_NBSN}
  (color online) Cation off-centerings along the Cartesian directions and
  their magnitude with respect to the center of their O cages (the 12 nearest
  O ions for the $A$-site and the 6 nearest for the $B$-site) in the relaxed
  40 atom NaBiScNbO$_6$ supercell.
}
\end{figure}

\begin{figure}[ht!]
\includegraphics[width=8cm,clip]{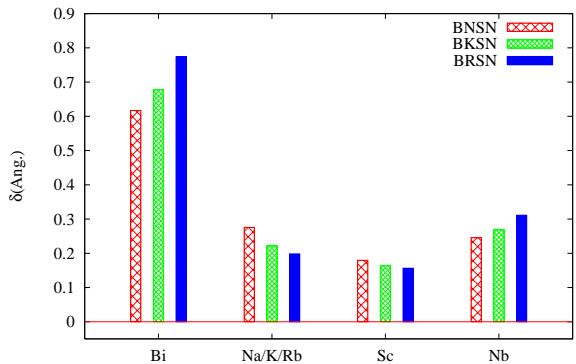}
\caption{\label{offcenter_BMSN}
  (color online) Average magnitude of cation off-centerings with respect to the
  center of their O cages in the relaxed 40 atom NaBiScNbO$_6$, KBiScNbO$_6$ and
  RbBiScNbO$_6$ supercells.
}
\end{figure}
  
  The cation off-centerings in NaBiScNbO$_6$ from the center of their
  O cages is shown in Fig. \ref{offcenter_NBSN}. All cations
  off-center along [111] direction and the magnitude of displacement
  is almost the same for a given cation. The Bi ions have a 
  large off-centering, which is a consequence of the difference in
  size compared to the other $A$-site ion and also its stereochemical
  activity. In addition to the off-centering of $A$-site cations,
  there are displacements of $B$-site cations as well. For
  NaBiScNbO$_6$, the average magnitudes of the cation off-centerings are
  0.62 \AA\ for Bi$^{3+}$, 0.28 \AA\ for Na$^{+}$, 0.18 \AA\ for
  Sc$^{3+}$ and 0.24 \AA\ for Nb$^{5+}$, respectively. This
  cooperation of all cations in the ferroelectric distortion is not
  unusual and is also seen in KNbO$_3$, BaTiO$_3$, PbTiO$_3$,
  BiPbZnNbO$_6$ and
  BiSrZnNbO$_6$.\cite{singh1992ferro,ghita2005prb,takagi2010prb} The
  cation off-centerings in KBiScNbO$_6$ and RbBiScNbO$_6$ are also
  collinear along [111] direction.

  As mentioned, the LDA lattice parameters of these supercells
  may be underestimated by 1-2$\%$. To study the robustness of
  ferroelectric displacements, we performed relaxation calculations
  for larger values of lattice parameters that lie between the equilibrium
  lattice parameters
(3.98 \AA, for NaBiScNbO$_6$, 4.03 \AA, for KBiScNbO$_6$ and 4.07
  \AA, for NaBiScNbO$_6$) and
the higher value of 4.23 \AA. We found ferroelectric cation off-centerings
  for each such calculation, confirming the robustness of ferroelectric
  distortions against LDA volume errors.

  Fig.~\ref{offcenter_BMSN} shows the average magnitudes of the cation
  off-centerings in NaBiScNbO$_6$, KBiScNbO$_6$ and RbBiScNbO$_6$.
  The displacements of Bi and Nb increase as the
  $A$-site size difference is increased. The reason for this increase
  in displacement is two-fold. First, the larger size difference
  decreases the tilting of $B$O$_6$ octahedra and provides more volume
  for displacement of Bi and Nb ions. Secondly, the lattice
  parameter increases as the size of $M$ ion increases. The
  ferroelectricity becomes stronger with increasing volume, as is
  usual in perovskite oxides. \cite{samara}
Competing with this is the fact that,
since polarization is dipole per unit volume, larger unit cell volume works
against high polarization. 
It is interesting to note that in
  contrast to Bi and Nb ions, the off-centerings become smaller for the
  $M$ ions, while there is no appreciable change in the displacement
  of Sc ion. As discussed below, there is little
  covalency between the unoccupied states of the $M$ ions with the occupied
  O 2$p$ states in these
materials. This works against
  off-centering.  Furthermore, $r_{\textrm{Na}^+}
  < r_{\textrm{K}+} < r_{\textrm{Rb}^+}$, and as the radius of an atom
  increases, there is less volume available for off-centering. This
  explains the smaller displacement of Rb ions as compared to Na ions.

\begin{figure}[ht!]
\includegraphics[width=9cm,clip]{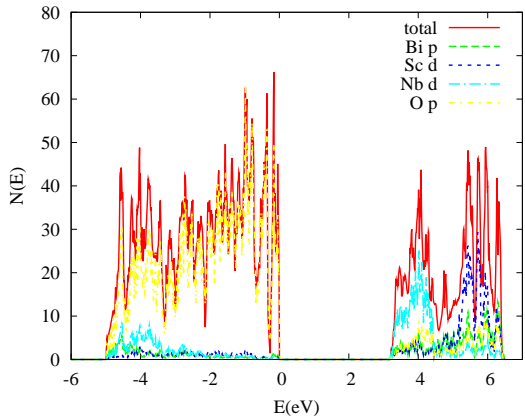}
\caption{\label{dos_NBSN}
  (color online) Electronic DOS of the pseudocubic relaxed NaBiScNbO$_6$ and the
  Bi $p$, Sc $d$, Nb $d$ and O $p$ projections. The valence band maximum is
  taken as the energy zero.
}
\end{figure}

  We now consider the electronic structure of these materials as it
  pertains to ferroelectricity. Fig. \ref{dos_NBSN} shows the electronic
  density of states (DOS) of the pseudocubic relaxed NaBiScNbO$_6$ and
  the projections onto Bi $p$, Sc $d$, Nb $d$ and O $p$
  pseudo-wavefunctions (the DOS of KBiScNbO$_6$ and RbBiScNbO$_6$ are
  almost identical and are not shown here).  The O 2$p$ states provide the
  DOS from -5 eV to the valence band maximum and there is a gap that
  separates the O 2$p$ derived states from the unoccupied metal
  states. The calculated LDA band gap is 3.2 eV, 3.1 eV and 3.1 eV for
  NaBiScNbO$_6$, KBiScNbO$_6$ and RbBiScNbO$_6$, respectively.
The actual band gaps are likely to be larger than these calculated ones
due to LDA band gap underestimation.
In any case, based on the calculated gaps,
  these materials are expected to be good insulators at ambient
  temperature.
  The O 2$p$ derived valence bands contain
  significant admixtures of
the nominally unoccupied metal states, which are mainly of Bi 6$p$, Sc
  3$d$ and Nb 4$d$ character. This mixing of occupied O 2$p$ states
  with unoccupied $d$ states of metals implies chemical bonding and is
  responsible for enhanced Born effective charges and ferroelectric
  off-centering, as in PbTiO$_3$ and
  BaTiO$_3$.\cite{cohen1992nature,zhong1994prl} Nb 4$d$ states are
  dominant near the conduction band minimum and this low energy favors
  hybridization with O 2$p$ states, as discussed in the case of
  PbMg$_{1/3}$Nb$_{2/3}$O$_3$ and KNbO$_3$ in comparison with
  KTaO$_3$.\cite{singh1996prb,suewattana2006prb} However, the Sc 3$d$
  states lie further above the conduction band minimum and this
  results in a somewhat weaker hybridization with O 2$p$ states.
  We do not find significant mixing of unoccupied Na/K/Rb $s$
  states with the occupied O 2$p$ states as may be expected
considering the very large electronegativity difference between these
elements and O.
  In NBSN, the influence of covalency is reflected in
  the enhanced Born effective charges of 4.38 for Bi, 4.73 for
  Sc and 6.42 for Nb as compared to the
very modest enhancement of (z$^*$=1.14) for Na.
  Similar numbers are obtained for KBSN and RBSN (see
  Table~\ref{tab:comparison}).
  
  In contrast to the trend in the cation off-centerings, the Born effective
  charges of Bi and Nb decrease, while those of $M$ and Sc
  ions increase when the $A$-site size difference becomes
  larger. These decreases in Born charges of Bi and Nb ions would mean
  lower values of polarization if the
Born charges were independent of atom position and
the off-centering were of the same magnitude
in all compounds. In fact, however, the polarizations as
  calculated with the Berry phase method are 73 $\mu$C/cm$^2$ for
  NBSN, 77 $\mu$C/cm$^2$ for KBSN, and 87 $\mu$C/cm$^2$. 

\begin{table}[ht!]
\caption{\label{tab:comparison}
  Comparison of properties of NaBiScNbO$_6$, KBiScNbO$_6$ and RbBiScNbO$_6$,
  where $r_{\text{M}^+}$ denotes the ion radii of $M$ ion, $a$
  denotes the pseudocubic lattice parameter, $z^*$ denotes the average Born
  effective charge, $\delta$ denotes the average cation off-centering with
  respect to the O cages, and $P$ is the magnitude of polarization.
}
\begin{ruledtabular}
\begin{tabular}{lccc}
               & NaBiScNbO$_6$    & KBiScNbO$_6$     & RbBiScNbO$_6$ \\ \hline
$r_{M^+}$(Na/K/Rb)
               & 1.53 \AA         & 1.78 \AA         & 1.86 \AA \\
$a$(LDA)       & 3.98 \AA         & 4.03 \AA         & 4.07 \AA \\
$\delta$(Bi)   & 0.62 \AA         & 0.67 \AA         & 0.77 \AA \\
$z^*$(Bi)      & 4.38             & 4.21             & 3.73 \\
$\delta$(Na/K/Rb)
               & 0.28 \AA         & 0.22 \AA         & 0.20 \AA \\
$z^*$(Na.K/Rb) & 1.14             & 1.17             & 1.22 \\
$\delta$(Sc)   & 0.18 \AA         & 0.16 \AA         & 0.16 \AA \\
$z^*$(Sc)      & 4.73             & 4.81             & 4.92 \\
$\delta$(Nb)   & 0.24 \AA         & 0.27 \AA         & 0.31 \AA \\
$z^*$(Nb)      & 6.42             & 6.23             & 5.80 \\
$P$(111)       & 73 $\mu$C/cm$^2$ & 77 $\mu$C/cm$^2$ & 87 $\mu$C/cm$^2$ \\
\end{tabular}
\end{ruledtabular}
\end{table}

  As mentioned, high piezoelectric performance is induced by
  the polarization rotation from the rhombohedral state to tetragonal
  state near the MPB. We studied the tendency towards tetragonality by
  calculating the total energies of NBSN, KBSN, and RBSN as a function
  of imposed tetragonal strain ($0.94<c/a<1.06$) while keeping the
  volumes equal to their equilibrium volume for the pseudocubic
  cells. In all materials, the lowest energy is at $c/a=1.00$ and the
  cation off-centerings are collinear and along [111] direction. The
  cations do not off center along [001] directions even with large $c/a$
and the
  off-centerings remain practically along the
[111] direction over the range of $c/a$
  values that we tried, implying robustness of the rhombohedral
  ferroelectricity. Therefore, these are rhombohedral ferroelectric
materials, and as such it is of interest to alloy with
  tetragonal ferroelectrics in order to produce a potential MPB.

  Related to this, a similar compound Na$_{0.5}$Bi$_{0.5}$TiO$_3$
  (NBT) is also a rhombohedral ferroelectric\cite{buhrer62} with $T_C
  = 593.15$ K and polarization 38 $\mu$C/cm$^2$.\cite{takenaka91}
  Solid solutions of NBT with tetragonal
  BaTiO$_3$ (NBBT) show a useful MPB with
  polarization 20 $\mu$C/cm$^2$.\cite{takenaka91} While the material
has not proved superior to other compounds
such as Pb(Zr,Ti)O$_3$ in transducers,
  it does have a lower free permittivity
  and a high electromechanical coupling factor that make it an interesting
  lead-free actuator material.
The materials in the present study are similar
  in that they are rhombohedral ferroelectrics and may be considered
to be related to NBT with Ti replaced by
  Sc$_{0.5}$Nb$_{0.5}$. By analogy, solid solutions of these materials
  with BaTiO$_3$ may provide materials that are useful as lead-free
  piezoelectrics. 
Another possibility may be alloying with the super tetragonal material
Bi(Zn,Ti)O$_3$. \cite{suchomel}

\section{CONCLUSIONS}

  We investigated the polar behavior of double perovskites
  NaBiScNbO$_6$, KBiScNbO$_6$ and RbBiScNbO$_6$ using density
  functional calculations on small supercells. We find high
  polarization along [111]
  directions. The main contribution to the polarization comes from
  large displacements of Bi and more modest displacements of Nb
ions, which, however, have high Born charges.
We find that the polarization
  increases as the $A$-site size difference becomes larger.  These
  materials also have substantial band gaps and the rhombohedral
  ferroelectricity is robust. 
 It will be of interest to investigate whether solid
  solutions of these materials with tetragonal ferroelectrics
  show MPBs possibly providing another avenue for development of lead-free
  piezoelectrics.

\section*{Acknowledgements}

  This work was supported by the Materials Sciences and
  Engineering Division, Office of Basic Energy Sciences,
U.S. Department of Energy
  (AS,VRC,DJS) and the Office of Naval Research (ST).

\bibliography{perovskite2}

\begin{thebibliography}{40}
\expandafter\ifx\csname natexlab\endcsname\relax\def\natexlab#1{#1}\fi
\expandafter\ifx\csname bibnamefont\endcsname\relax
  \def\bibnamefont#1{#1}\fi
\expandafter\ifx\csname bibfnamefont\endcsname\relax
  \def\bibfnamefont#1{#1}\fi
\expandafter\ifx\csname citenamefont\endcsname\relax
  \def\citenamefont#1{#1}\fi
\expandafter\ifx\csname url\endcsname\relax
  \def\url#1{\texttt{#1}}\fi
\expandafter\ifx\csname urlprefix\endcsname\relax\def\urlprefix{URL }\fi
\providecommand{\bibinfo}[2]{#2}
\providecommand{\eprint}[2][]{\url{#2}}

\bibitem[{\citenamefont{Scott}(2007)}]{scott}
\bibinfo{author}{\bibfnamefont{J.~F.} \bibnamefont{Scott}},
  \bibinfo{journal}{Science} \textbf{\bibinfo{volume}{315}},
  \bibinfo{pages}{954} (\bibinfo{year}{2007}).

\bibitem[{\citenamefont{Ahn et~al.}(2004)\citenamefont{Ahn, Rabe, and
  Triscone}}]{ahn}
\bibinfo{author}{\bibfnamefont{C.~H.} \bibnamefont{Ahn}},
  \bibinfo{author}{\bibfnamefont{K.~M.} \bibnamefont{Rabe}}, \bibnamefont{and}
  \bibinfo{author}{\bibfnamefont{J.~M.} \bibnamefont{Triscone}},
  \bibinfo{journal}{Science} \textbf{\bibinfo{volume}{303}},
  \bibinfo{pages}{488} (\bibinfo{year}{2004}).

\bibitem[{\citenamefont{Lee et~al.}(2005)\citenamefont{Lee, Christen, Chisholm,
  Rouleau, and Lowndes}}]{lee}
\bibinfo{author}{\bibfnamefont{H.~N.} \bibnamefont{Lee}},
  \bibinfo{author}{\bibfnamefont{H.~M.} \bibnamefont{Christen}},
  \bibinfo{author}{\bibfnamefont{M.~F.} \bibnamefont{Chisholm}},
  \bibinfo{author}{\bibfnamefont{C.~M.} \bibnamefont{Rouleau}},
  \bibnamefont{and} \bibinfo{author}{\bibfnamefont{D.~H.}
  \bibnamefont{Lowndes}}, \bibinfo{journal}{Nature (London)}
  \textbf{\bibinfo{volume}{433}}, \bibinfo{pages}{395} (\bibinfo{year}{2005}).

\bibitem[{\citenamefont{Dawber et~al.}(2005)\citenamefont{Dawber, Rabe, and
  Scott}}]{dawber}
\bibinfo{author}{\bibfnamefont{M.}~\bibnamefont{Dawber}},
  \bibinfo{author}{\bibfnamefont{K.~M.} \bibnamefont{Rabe}}, \bibnamefont{and}
  \bibinfo{author}{\bibfnamefont{J.~F.} \bibnamefont{Scott}},
  \bibinfo{journal}{Rev. Mod. Phys.} \textbf{\bibinfo{volume}{77}},
  \bibinfo{pages}{1083} (\bibinfo{year}{2005}).

\bibitem[{\citenamefont{Cohen and Krakauer}(1990)}]{cohen1990prb}
\bibinfo{author}{\bibfnamefont{R.~E.} \bibnamefont{Cohen}} \bibnamefont{and}
  \bibinfo{author}{\bibfnamefont{H.}~\bibnamefont{Krakauer}},
  \bibinfo{journal}{Phys. Rev. B} \textbf{\bibinfo{volume}{42}},
  \bibinfo{pages}{6416} (\bibinfo{year}{1990}).

\bibitem[{\citenamefont{Cohen}(1992)}]{cohen1992nature}
\bibinfo{author}{\bibfnamefont{R.~E.} \bibnamefont{Cohen}},
  \bibinfo{journal}{Nature} \textbf{\bibinfo{volume}{358}},
  \bibinfo{pages}{136} (\bibinfo{year}{1992}).

\bibitem[{\citenamefont{Cooper and Rabe}(2009)}]{cooper}
\bibinfo{author}{\bibfnamefont{V.~R.} \bibnamefont{Cooper}} \bibnamefont{and}
  \bibinfo{author}{\bibfnamefont{K.~M.} \bibnamefont{Rabe}},
  \bibinfo{journal}{Phys. Rev. B} \textbf{\bibinfo{volume}{79}},
  \bibinfo{pages}{180101} (\bibinfo{year}{2009}).

\bibitem[{\citenamefont{Park and Shrout}(1997)}]{park1997jap}
\bibinfo{author}{\bibfnamefont{S.-E.} \bibnamefont{Park}} \bibnamefont{and}
  \bibinfo{author}{\bibfnamefont{T.~R.} \bibnamefont{Shrout}},
  \bibinfo{journal}{J. Appl. Phys.} \textbf{\bibinfo{volume}{82}},
  \bibinfo{pages}{1804} (\bibinfo{year}{1997}).

\bibitem[{\citenamefont{Noheda et~al.}(1999)\citenamefont{Noheda, Cox, Shirane,
  Cross, and Zhong}}]{noheda1999apl}
\bibinfo{author}{\bibfnamefont{B.}~\bibnamefont{Noheda}},
  \bibinfo{author}{\bibfnamefont{D.~E.} \bibnamefont{Cox}},
  \bibinfo{author}{\bibfnamefont{G.}~\bibnamefont{Shirane}},
  \bibinfo{author}{\bibfnamefont{L.~E.} \bibnamefont{Cross}}, \bibnamefont{and}
  \bibinfo{author}{\bibfnamefont{Z.}~\bibnamefont{Zhong}},
  \bibinfo{journal}{Appl. Phys. Lett.} \textbf{\bibinfo{volume}{74}},
  \bibinfo{pages}{2059} (\bibinfo{year}{1999}).

\bibitem[{\citenamefont{Bellaiche et~al.}(2000)\citenamefont{Bellaiche, Garcia,
  and Vanderbilt}}]{bellaiche2000prl}
\bibinfo{author}{\bibfnamefont{L.}~\bibnamefont{Bellaiche}},
  \bibinfo{author}{\bibfnamefont{A.}~\bibnamefont{Garcia}}, \bibnamefont{and}
  \bibinfo{author}{\bibfnamefont{D.}~\bibnamefont{Vanderbilt}},
  \bibinfo{journal}{Phys. Rev. Lett.} \textbf{\bibinfo{volume}{84}},
  \bibinfo{pages}{5427} (\bibinfo{year}{2000}).

\bibitem[{\citenamefont{Fu and Cohen}(2000)}]{fu2000nature}
\bibinfo{author}{\bibfnamefont{H.}~\bibnamefont{Fu}} \bibnamefont{and}
  \bibinfo{author}{\bibfnamefont{R.~E.} \bibnamefont{Cohen}},
  \bibinfo{journal}{Nature} \textbf{\bibinfo{volume}{403}},
  \bibinfo{pages}{281} (\bibinfo{year}{2000}).

\bibitem[{\citenamefont{Guo et~al.}(2000)\citenamefont{Guo, Cross, Park,
  Noheda, Cox, and Shirane}}]{guo2000prl}
\bibinfo{author}{\bibfnamefont{R.}~\bibnamefont{Guo}},
  \bibinfo{author}{\bibfnamefont{L.~E.} \bibnamefont{Cross}},
  \bibinfo{author}{\bibfnamefont{S.-E.} \bibnamefont{Park}},
  \bibinfo{author}{\bibfnamefont{B.}~\bibnamefont{Noheda}},
  \bibinfo{author}{\bibfnamefont{D.~E.} \bibnamefont{Cox}}, \bibnamefont{and}
  \bibinfo{author}{\bibfnamefont{G.}~\bibnamefont{Shirane}},
  \bibinfo{journal}{Phys. Rev. Lett.} \textbf{\bibinfo{volume}{84}},
  \bibinfo{pages}{5423} (\bibinfo{year}{2000}).

\bibitem[{\citenamefont{Noheda et~al.}(2001)\citenamefont{Noheda, Cox, Shirane,
  Park, Cross, and Zhong}}]{noheda2001prl}
\bibinfo{author}{\bibfnamefont{B.}~\bibnamefont{Noheda}},
  \bibinfo{author}{\bibfnamefont{D.~E.} \bibnamefont{Cox}},
  \bibinfo{author}{\bibfnamefont{G.}~\bibnamefont{Shirane}},
  \bibinfo{author}{\bibfnamefont{S.-E.} \bibnamefont{Park}},
  \bibinfo{author}{\bibfnamefont{L.~E.} \bibnamefont{Cross}}, \bibnamefont{and}
  \bibinfo{author}{\bibfnamefont{Z.}~\bibnamefont{Zhong}},
  \bibinfo{journal}{Phys. Rev. Lett.} \textbf{\bibinfo{volume}{86}},
  \bibinfo{pages}{3891} (\bibinfo{year}{2001}).

\bibitem[{\citenamefont{Singh and Park}(2008)}]{singh2008prl}
\bibinfo{author}{\bibfnamefont{D.~J.} \bibnamefont{Singh}} \bibnamefont{and}
  \bibinfo{author}{\bibfnamefont{C.~H.} \bibnamefont{Park}},
  \bibinfo{journal}{Phys. Rev. Lett.} \textbf{\bibinfo{volume}{100}},
  \bibinfo{pages}{087601} (\bibinfo{year}{2008}).

\bibitem[{\citenamefont{Fu et~al.}(2008)\citenamefont{Fu, Itoh, Koshihara,
  Kosugi, and Tsuneyuki}}]{desheng2008prl}
\bibinfo{author}{\bibfnamefont{D.}~\bibnamefont{Fu}},
  \bibinfo{author}{\bibfnamefont{M.}~\bibnamefont{Itoh}},
  \bibinfo{author}{\bibfnamefont{S.~Y.} \bibnamefont{Koshihara}},
  \bibinfo{author}{\bibfnamefont{T.}~\bibnamefont{Kosugi}}, \bibnamefont{and}
  \bibinfo{author}{\bibfnamefont{S.}~\bibnamefont{Tsuneyuki}},
  \bibinfo{journal}{Phys. Rev. Lett.} \textbf{\bibinfo{volume}{100}},
  \bibinfo{pages}{227601} (\bibinfo{year}{2008}).

\bibitem[{\citenamefont{Kim et~al.}(2010)\citenamefont{Kim, Christen,
  Biegalski, and Singh}}]{bcz}
\bibinfo{author}{\bibfnamefont{H.~S.} \bibnamefont{Kim}},
  \bibinfo{author}{\bibfnamefont{H.~M.} \bibnamefont{Christen}},
  \bibinfo{author}{\bibfnamefont{M.~D.} \bibnamefont{Biegalski}},
  \bibnamefont{and} \bibinfo{author}{\bibfnamefont{D.~J.} \bibnamefont{Singh}},
  \bibinfo{journal}{J. Appl. Phys.} \textbf{\bibinfo{volume}{108}},
  \bibinfo{pages}{in press} (\bibinfo{year}{2010}).

\bibitem[{\citenamefont{Takagi et~al.}(2010)\citenamefont{Takagi, Subedi,
  Singh, and Cooper}}]{takagi2010prb}
\bibinfo{author}{\bibfnamefont{S.}~\bibnamefont{Takagi}},
  \bibinfo{author}{\bibfnamefont{A.}~\bibnamefont{Subedi}},
  \bibinfo{author}{\bibfnamefont{D.~J.} \bibnamefont{Singh}}, \bibnamefont{and}
  \bibinfo{author}{\bibfnamefont{V.~R.} \bibnamefont{Cooper}},
  \bibinfo{journal}{Phys. Rev. B} \textbf{\bibinfo{volume}{81}},
  \bibinfo{pages}{134106} (\bibinfo{year}{2010}).

\bibitem[{\citenamefont{Neaton et~al.}(2005)\citenamefont{Neaton, Ederer,
  Waghmare, Spaldin, and Rabe}}]{neaton2005prb}
\bibinfo{author}{\bibfnamefont{J.~B.} \bibnamefont{Neaton}},
  \bibinfo{author}{\bibfnamefont{C.}~\bibnamefont{Ederer}},
  \bibinfo{author}{\bibfnamefont{U.~V.} \bibnamefont{Waghmare}},
  \bibinfo{author}{\bibfnamefont{N.~A.} \bibnamefont{Spaldin}},
  \bibnamefont{and} \bibinfo{author}{\bibfnamefont{K.~M.} \bibnamefont{Rabe}},
  \bibinfo{journal}{Phys. Rev. B} \textbf{\bibinfo{volume}{71}},
  \bibinfo{pages}{014113} (\bibinfo{year}{2005}).

\bibitem[{\citenamefont{Eitel et~al.}(2004)\citenamefont{Eitel, Zhang, Shrout,
  Randall, and Levin}}]{eitel}
\bibinfo{author}{\bibfnamefont{R.~E.} \bibnamefont{Eitel}},
  \bibinfo{author}{\bibfnamefont{S.~J.} \bibnamefont{Zhang}},
  \bibinfo{author}{\bibfnamefont{T.~R.} \bibnamefont{Shrout}},
  \bibinfo{author}{\bibfnamefont{C.~A.} \bibnamefont{Randall}},
  \bibnamefont{and} \bibinfo{author}{\bibfnamefont{I.}~\bibnamefont{Levin}},
  \bibinfo{journal}{J. Appl. Phys.} \textbf{\bibinfo{volume}{96}},
  \bibinfo{pages}{2828} (\bibinfo{year}{2004}).

\bibitem[{\citenamefont{Iniguez et~al.}(2003)\citenamefont{Iniguez, Vanderbilt,
  and Bellaiche}}]{iniguez}
\bibinfo{author}{\bibfnamefont{J.}~\bibnamefont{Iniguez}},
  \bibinfo{author}{\bibfnamefont{D.}~\bibnamefont{Vanderbilt}},
  \bibnamefont{and}
  \bibinfo{author}{\bibfnamefont{L.}~\bibnamefont{Bellaiche}},
  \bibinfo{journal}{Phys. Rev. B} \textbf{\bibinfo{volume}{67}},
  \bibinfo{pages}{224107} (\bibinfo{year}{2003}).

\bibitem[{\citenamefont{Halilov et~al.}(2004)\citenamefont{Halilov, Fornari,
  and Singh}}]{halilov2004prb}
\bibinfo{author}{\bibfnamefont{S.~V.} \bibnamefont{Halilov}},
  \bibinfo{author}{\bibfnamefont{M.}~\bibnamefont{Fornari}}, \bibnamefont{and}
  \bibinfo{author}{\bibfnamefont{D.~J.} \bibnamefont{Singh}},
  \bibinfo{journal}{Phys. Rev. B} \textbf{\bibinfo{volume}{69}},
  \bibinfo{pages}{174107} (\bibinfo{year}{2004}).

\bibitem[{\citenamefont{Baroni et~al.}()\citenamefont{Baroni, dal Corso,
  de~Gironcoli, Giannozzi, Cavazzoni, Ballabio, Scandolo, Chiarotti, Focher,
  Pasquarello et~al.}}]{espresso}
\bibinfo{author}{\bibfnamefont{S.}~\bibnamefont{Baroni}},
  \bibinfo{author}{\bibfnamefont{A.}~\bibnamefont{dal Corso}},
  \bibinfo{author}{\bibfnamefont{S.}~\bibnamefont{de~Gironcoli}},
  \bibinfo{author}{\bibfnamefont{P.}~\bibnamefont{Giannozzi}},
  \bibinfo{author}{\bibfnamefont{C.}~\bibnamefont{Cavazzoni}},
  \bibinfo{author}{\bibfnamefont{G.}~\bibnamefont{Ballabio}},
  \bibinfo{author}{\bibfnamefont{S.}~\bibnamefont{Scandolo}},
  \bibinfo{author}{\bibfnamefont{G.}~\bibnamefont{Chiarotti}},
  \bibinfo{author}{\bibfnamefont{P.}~\bibnamefont{Focher}},
  \bibinfo{author}{\bibfnamefont{A.}~\bibnamefont{Pasquarello}},
  \bibnamefont{et~al.}, \bibinfo{howpublished}{URL
  http://www.quantum-espresso.org}.

\bibitem[{\citenamefont{Vanderbilt}(1990)}]{uspp}
\bibinfo{author}{\bibfnamefont{D.}~\bibnamefont{Vanderbilt}},
  \bibinfo{journal}{Phys. Rev. B} \textbf{\bibinfo{volume}{41}},
  \bibinfo{pages}{7892} (\bibinfo{year}{1990}).

\bibitem[{\citenamefont{Chu et~al.}(1995)\citenamefont{Chu, Reaney, and
  Setter}}]{chu1995jap}
\bibinfo{author}{\bibfnamefont{F.}~\bibnamefont{Chu}},
  \bibinfo{author}{\bibfnamefont{I.~M.} \bibnamefont{Reaney}},
  \bibnamefont{and} \bibinfo{author}{\bibfnamefont{N.}~\bibnamefont{Setter}},
  \bibinfo{journal}{J. Appl. Phys} \textbf{\bibinfo{volume}{77}},
  \bibinfo{pages}{1671} (\bibinfo{year}{1995}).

\bibitem[{\citenamefont{Malibert et~al.}(1997)\citenamefont{Malibert, Dkhil,
  Kiat, Durand, B{\'{e}}rar, and Spasojevic-de{\
  }Bir{\'{e}}}}]{malibert1997jpcond}
\bibinfo{author}{\bibfnamefont{C.}~\bibnamefont{Malibert}},
  \bibinfo{author}{\bibfnamefont{B.}~\bibnamefont{Dkhil}},
  \bibinfo{author}{\bibfnamefont{J.~M.} \bibnamefont{Kiat}},
  \bibinfo{author}{\bibfnamefont{D.}~\bibnamefont{Durand}},
  \bibinfo{author}{\bibfnamefont{J.~F.} \bibnamefont{B{\'{e}}rar}},
  \bibnamefont{and}
  \bibinfo{author}{\bibfnamefont{A.}~\bibnamefont{Spasojevic-de{\
  }Bir{\'{e}}}}, \bibinfo{journal}{J. Phys.: Condens. Matter}
  \textbf{\bibinfo{volume}{9}}, \bibinfo{pages}{7485} (\bibinfo{year}{1997}).

\bibitem[{\citenamefont{Park et~al.}(1994)\citenamefont{Park, Chung, Kim, and
  Hong}}]{park1994jamceram}
\bibinfo{author}{\bibfnamefont{S.-E.} \bibnamefont{Park}},
  \bibinfo{author}{\bibfnamefont{S.-J.} \bibnamefont{Chung}},
  \bibinfo{author}{\bibfnamefont{I.-T.} \bibnamefont{Kim}}, \bibnamefont{and}
  \bibinfo{author}{\bibfnamefont{K.~S.} \bibnamefont{Hong}},
  \bibinfo{journal}{J. Am. Ceram. Soc.} \textbf{\bibinfo{volume}{77}},
  \bibinfo{pages}{2641} (\bibinfo{year}{1994}).

\bibitem[{\citenamefont{Jones et~al.}(2002)\citenamefont{Jones, Kreisel, and
  Thomas}}]{jones2002powder}
\bibinfo{author}{\bibfnamefont{G.~O.} \bibnamefont{Jones}},
  \bibinfo{author}{\bibfnamefont{J.}~\bibnamefont{Kreisel}}, \bibnamefont{and}
  \bibinfo{author}{\bibfnamefont{P.~A.} \bibnamefont{Thomas}},
  \bibinfo{journal}{Powder Diffr.} \textbf{\bibinfo{volume}{17}},
  \bibinfo{pages}{301} (\bibinfo{year}{2002}).

\bibitem[{\citenamefont{Jones and Thomas}(2002)}]{jones2002acta}
\bibinfo{author}{\bibfnamefont{G.~O.} \bibnamefont{Jones}} \bibnamefont{and}
  \bibinfo{author}{\bibfnamefont{P.~A.} \bibnamefont{Thomas}},
  \bibinfo{journal}{Acta Crystallogr. B} \textbf{\bibinfo{volume}{58}},
  \bibinfo{pages}{168} (\bibinfo{year}{2002}).

\bibitem[{\citenamefont{Petzelt et~al.}(2004)\citenamefont{Petzelt, Kamba,
  Febry, Noujni, Porokhonskyy, Pashkin, Franke, Rolder, Suchnicz, Klein
  et~al.}}]{petzelt2004jphys}
\bibinfo{author}{\bibfnamefont{J.}~\bibnamefont{Petzelt}},
  \bibinfo{author}{\bibfnamefont{S.}~\bibnamefont{Kamba}},
  \bibinfo{author}{\bibfnamefont{J.}~\bibnamefont{Febry}},
  \bibinfo{author}{\bibfnamefont{D.}~\bibnamefont{Noujni}},
  \bibinfo{author}{\bibfnamefont{V.}~\bibnamefont{Porokhonskyy}},
  \bibinfo{author}{\bibfnamefont{A.}~\bibnamefont{Pashkin}},
  \bibinfo{author}{\bibfnamefont{I.}~\bibnamefont{Franke}},
  \bibinfo{author}{\bibfnamefont{K.}~\bibnamefont{Rolder}},
  \bibinfo{author}{\bibfnamefont{J.}~\bibnamefont{Suchnicz}},
  \bibinfo{author}{\bibfnamefont{R.}~\bibnamefont{Klein}},
  \bibnamefont{et~al.}, \bibinfo{journal}{J. Phys.: Condens. Matter}
  \textbf{\bibinfo{volume}{16}}, \bibinfo{pages}{2719} (\bibinfo{year}{2004}).

\bibitem[{\citenamefont{Glazer}(1972)}]{glazer1972acta}
\bibinfo{author}{\bibfnamefont{A.~M.} \bibnamefont{Glazer}},
  \bibinfo{journal}{Acta Crystallogr. Sect. B} \textbf{\bibinfo{volume}{28}},
  \bibinfo{pages}{3384} (\bibinfo{year}{1972}).

\bibitem[{\citenamefont{Fornari and Singh}(2001)}]{fornari-tilt}
\bibinfo{author}{\bibfnamefont{M.}~\bibnamefont{Fornari}} \bibnamefont{and}
  \bibinfo{author}{\bibfnamefont{D.~J.} \bibnamefont{Singh}},
  \bibinfo{journal}{Phys. Rev. B} \textbf{\bibinfo{volume}{63}},
  \bibinfo{pages}{092101} (\bibinfo{year}{2001}).

\bibitem[{\citenamefont{Singh and Boyer}(1992)}]{singh1992ferro}
\bibinfo{author}{\bibfnamefont{D.~J.} \bibnamefont{Singh}} \bibnamefont{and}
  \bibinfo{author}{\bibfnamefont{L.~L.} \bibnamefont{Boyer}},
  \bibinfo{journal}{Ferroelectrics} \textbf{\bibinfo{volume}{136}},
  \bibinfo{pages}{95} (\bibinfo{year}{1992}).

\bibitem[{\citenamefont{Ghita et~al.}(2005)\citenamefont{Ghita, Fornari, Singh,
  and Halilov}}]{ghita2005prb}
\bibinfo{author}{\bibfnamefont{M.}~\bibnamefont{Ghita}},
  \bibinfo{author}{\bibfnamefont{M.}~\bibnamefont{Fornari}},
  \bibinfo{author}{\bibfnamefont{D.~J.} \bibnamefont{Singh}}, \bibnamefont{and}
  \bibinfo{author}{\bibfnamefont{S.~V.} \bibnamefont{Halilov}},
  \bibinfo{journal}{Phys. Rev. B} \textbf{\bibinfo{volume}{72}},
  \bibinfo{pages}{054114} (\bibinfo{year}{2005}).

\bibitem[{\citenamefont{Samara and Peercy}(1981)}]{samara}
\bibinfo{author}{\bibfnamefont{G.}~\bibnamefont{Samara}} \bibnamefont{and}
  \bibinfo{author}{\bibfnamefont{P.}~\bibnamefont{Peercy}},
  \bibinfo{journal}{Solid State Physics-Advances In Research And Applications}
  \textbf{\bibinfo{volume}{36}}, \bibinfo{pages}{1} (\bibinfo{year}{1981}).

\bibitem[{\citenamefont{Zhong et~al.}(1994)\citenamefont{Zhong, King-Smith, and
  Vanderbilt}}]{zhong1994prl}
\bibinfo{author}{\bibfnamefont{W.}~\bibnamefont{Zhong}},
  \bibinfo{author}{\bibfnamefont{R.~D.} \bibnamefont{King-Smith}},
  \bibnamefont{and}
  \bibinfo{author}{\bibfnamefont{D.}~\bibnamefont{Vanderbilt}},
  \bibinfo{journal}{Phys. Rev. Lett.} \textbf{\bibinfo{volume}{72}},
  \bibinfo{pages}{3618} (\bibinfo{year}{1994}).

\bibitem[{\citenamefont{Singh}(1996)}]{singh1996prb}
\bibinfo{author}{\bibfnamefont{D.~J.} \bibnamefont{Singh}},
  \bibinfo{journal}{Phys. Rev. B} \textbf{\bibinfo{volume}{53}},
  \bibinfo{pages}{176} (\bibinfo{year}{1996}).

\bibitem[{\citenamefont{Suewattana and Singh}(2006)}]{suewattana2006prb}
\bibinfo{author}{\bibfnamefont{M.}~\bibnamefont{Suewattana}} \bibnamefont{and}
  \bibinfo{author}{\bibfnamefont{D.~J.} \bibnamefont{Singh}},
  \bibinfo{journal}{Phys. Rev. B} \textbf{\bibinfo{volume}{73}},
  \bibinfo{pages}{224105} (\bibinfo{year}{2006}).

\bibitem[{\citenamefont{Buhrer}(1962)}]{buhrer62}
\bibinfo{author}{\bibfnamefont{C.~F.} \bibnamefont{Buhrer}},
  \bibinfo{journal}{J. Chem. Phys.} \textbf{\bibinfo{volume}{36}},
  \bibinfo{pages}{798} (\bibinfo{year}{1962}).

\bibitem[{\citenamefont{Takenaka et~al.}(1991)\citenamefont{Takenaka, Maruyama,
  and Sakata}}]{takenaka91}
\bibinfo{author}{\bibfnamefont{T.}~\bibnamefont{Takenaka}},
  \bibinfo{author}{\bibfnamefont{K.}~\bibnamefont{Maruyama}}, \bibnamefont{and}
  \bibinfo{author}{\bibfnamefont{K.}~\bibnamefont{Sakata}},
  \bibinfo{journal}{Jpn. J. Appl. Phys.} \textbf{\bibinfo{volume}{30}},
  \bibinfo{pages}{2236} (\bibinfo{year}{1991}).

\bibitem[{\citenamefont{Suchomel et~al.}(2006)\citenamefont{Suchomel, Fogg,
  Allix, Niu, Claridge, and Rosseinsky}}]{suchomel}
\bibinfo{author}{\bibfnamefont{M.~R.} \bibnamefont{Suchomel}},
  \bibinfo{author}{\bibfnamefont{A.~M.} \bibnamefont{Fogg}},
  \bibinfo{author}{\bibfnamefont{M.}~\bibnamefont{Allix}},
  \bibinfo{author}{\bibfnamefont{H.~J.} \bibnamefont{Niu}},
  \bibinfo{author}{\bibfnamefont{J.~B.} \bibnamefont{Claridge}},
  \bibnamefont{and} \bibinfo{author}{\bibfnamefont{M.~J.}
  \bibnamefont{Rosseinsky}}, \bibinfo{journal}{Chem. Mater.}
  \textbf{\bibinfo{volume}{18}}, \bibinfo{pages}{4987} (\bibinfo{year}{2006}).

\end{thebibliography}

\end{document}